\documentclass[%
 aip,
 amsmath,amssymb,
 reprint,%
]{revtex4-1}

\usepackage{graphicx}% Include figure files
\usepackage{dcolumn}% Align table columns on decimal point
\usepackage{bm}% bold math
\usepackage[utf8]{inputenc}
\usepackage[T1]{fontenc}
\usepackage{mathptmx}
\usepackage{etoolbox}

\usepackage{comment}

\newcommand{\bu}{{\bf u}}

\newcommand{\bx}{{\bf x}}
\newcommand{\bn}{{\bf n}}
\newcommand{\bnh}{\hat{\bn}}

\newcommand{\bV}{{\bf V}}
\newcommand{\bomega}{{\mbox{\boldmath $\omega$}}}

\newcommand{\be}{\begin{equation}}
\newcommand{\ee}{\end{equation}} 
\newcommand{\lb}{\label}

\newcommand{\cS}{{\mathcal S}}
\newcommand{\cH}{{\mathcal H}}
\newcommand{\cD}{{\mathcal D}}

\newcommand{\grad}{{\mbox{\boldmath $\nabla$}}}

\newcommand{\bdot}{{\mbox{\boldmath $\cdot$}}}

\newcommand{\btimes}{{\mbox{\boldmath $\times$}}}

\makeatletter
\def\@email#1#2{%
 \endgroup
 \patchcmd{\titleblock@produce}
  {\frontmatter@RRAPformat}
  {\frontmatter@RRAPformat{\produce@RRAP{*#1\href{mailto:#2}{#2}}}\frontmatter@RRAPformat}
  {}{}
}%
\makeatother
\begin{document}

\preprint{AIP/123-QED}

\title[On Galilean Invariance of Mean Kinetic Helicity]{}

\author{D. Soltani Tehrani}
\affiliation{Mechanical Engineering Department, University of Rochester, Rochester, NY 14625, USA}
\email{dsoltani@ur.rochester.edu.}

\author{H. Aluie}
\affiliation{Mechanical Engineering Department, University of Rochester, Rochester, NY 14625, USA}
\affiliation{Laboratory for Laser Energetics, University of Rochester, Rochester, NY 14625, USA.}
%\email{hussein@rochester.edu.}

\date{\today}% It is always \today, today,
             %  but any date may be explicitly specified

\begin{abstract}
While kinetic helicity is not Galilean invariant locally, it is known (K. Moffatt, Journal of Fluid Mechanics, 35, 117 (1969)) that its spatial integral quantifies the degree of knottedness of vorticity field lines. Being a topological property of the flow, mean kinetic helicity is Galilean invariant. Here, we provide a direct mathematical proof that kinetic helicity is Galilean invariant when spatially integrated over regions enclosed by vorticity surfaces, \textit{i.e.}, surfaces of zero vorticity flux. We also discuss so-called ``relative'' kinetic helicity, which is Galilean invariant when integrated over any region in the flow. 
\end{abstract}

\maketitle

Kinetic helicity is a quadratic quantity that was shown to be a global invariant in ideal fluids by Moreau in 1961 \cite{moreau1961constants}. Its magnetic analogue was discovered earlier by Els\"asser\cite{elsasser1956hydromagnetic}  in 1956 and, independently, by Woltjer\cite{woltjer1958theorem} in 1958. The topological significance of helicity was first recognized by Moffatt\cite{moffatt1969degree} in 1969, who showed that it quantifies the degree of knottedness of vorticity (or magnetic) field lines in a system. This topological interpretation was later proved in a more general setting by Arnold \cite{Arnold1973,arnold1992topological}.

Local kinetic helicity is defined as the projection of vorticity along the velocity direction, $h(\bx) = \bu \cdot \bomega$, where $\bu$ is the flow velocity and $\bomega=\grad\btimes\bu$ is vorticity. Helicity quantifies the local right- or left-handedness of helical streamlines, corresponding to positive or negative local kinetic helicity. It is well-known that it plays an important role in the nonlinear evolution of flows \cite{pouquet2019helicity,angriman2021broken}. It is evident that local helicity is not Galilean invariant \cite{speziale1987helicity,moffatt1992helicity,pope2000turbulent}.

Mean kinetic helicity, being a volume integral of $h(\bx)$, is a global (system-wide) measure of handedness or the statistical lack of reflection symmetry in a flow. Since mean kinetic helicity characterizes a topological property of the flow, it is Galilean invariant \cite{moffatt1992helicity}. Mean helicity has been shown to play an important role in the cascade of energy in turbulent flows \cite{biferale2012inverse,baj2022simultaneous} and can lead to large-scale instabilities and mean flow generation \cite{frisch1987large,inagaki2017mechanism}. In geophysical and astrophysical settings, strong helicity can hinder the downscale cascade process and the associated dissipation of energy \cite{teitelbaum2011decay,rorai2013helicity,buzzicotti2018energy}.

Here, we present a direct mathematical proof that mean helicity over regions enclosed by vorticity surfaces is Galilean invariant without appealing to its topological meaning. We then show that so-called ``relative'' kinetic helicity is Galilean invariant over any region.
Showing that helicity is Galilean invariant is important for practical modeling considerations because it justifies the usage of simulations of helical flows in idealized domains (e.g., a periodic box) as representative of small regions embedded in much larger systems (e.g., a hurricane). Galilean invariance indicates that sweeping due to flow at the global scale should not alter the representation of helicity in local regions. Galilean invariance is also important for practical measurement considerations since it is often the case that we are only able to measure a flow in local regions in a system, such as the solar wind.

% \section{Proof}
Mean (or spatially-averaged) kinetic helicity is 
\begin{eqnarray}
    \mathcal{H} &=& \int_{\mathcal{D}}^{} d^3\bx ~ \bu \cdot \bomega,
    \label{eq:meanHelicity}
\end{eqnarray}
where $d^3\bx$ is a volume element, $\bu$ is the flow velocity of an inviscid incompressible fluid and $\bomega=\grad\btimes\bu$ is its associated vorticity. $\mathcal{D}$ is any region in the flow domain enclosed by a vorticity surface $\mathcal{S}$. In analogy with a magnetic surface \cite{berger1984topological}, a vorticity surface $\cS$ has zero vorticity flux, $\bomega \cdot \bnh = 0$, where $\bnh$ is the local normal vector to $\cS$.

Consider two frames of reference, $F: {\bx}, t$ and $F^{'}: {\bx}^{'}, t^{'},$ related by a Galilean transformation
\begin{subequations}
\label{eq:GT}
\begin{align}
    {\bx}^{'} &= \bx + \bV t + \bx_0 ~,\label{eq:GT_x}\\
    t^{'} &= t ~,\label{eq:GT_t}
\end{align}
\end{subequations}
where $\bx_0$ is a constant translation and $\bV$ is a constant velocity boost. From eq.~\eqref{eq:GT_x}, we have that the flow velocity in frame $F'$ is
\begin{eqnarray}
    {\bu}^{'} &=& \bu + \bV~.
    \label{eq:GT_u}
\end{eqnarray}
Spatial derivatives of the flow, including vorticity, are invariant under Galilean transformations,
\begin{eqnarray}
    \frac{\partial {u_i}^{'}}{\partial {x_j}^{'}} &=& \frac{\partial {u_i}}{{\partial x_j}}~.
    \label{eq:GT_grad}
\end{eqnarray}

We will now prove that mean kinetic helicity $\cH$ is Galilean invariant,  ${\mathcal{H}}^{'} = \mathcal{H}$. In frame $F'$, we have 
\begin{subequations}
\lb{eq:GT_H}
\begin{align}
    {\mathcal{H}}^{'} &= \int_{\mathcal{D}}^{} d^3\bx ~ {\bu}^{'} \cdot {\bomega}^{'} ~ \lb{eq:GT_H_1}\\
%    &= \int_{\mathcal{D}}^{} d^3\bx ~ (\bu - \bV) \cdot {\bomega} ~ \\
    &= {\int_{\mathcal{D}}^{} d^3\bx ~ \bu \cdot {\bomega} ~ + \int_{\mathcal{D}}^{} d^3\bx ~ \bV \cdot {\bomega} ~} \lb{eq:GT_H_2}\\
    &= \hspace{.4cm}{\mathcal{H} \hspace{1.08cm}+ \int_{\mathcal{D}}^{} d^3\bx ~ \bV \cdot {\bomega} ~.}
    \lb{eq:GT_H_3}
\end{align}
\end{subequations}

Since $\bV$ is a constant velocity, it can be written as the gradient of a  potential field,
\begin{eqnarray}
    \bV &=& \grad \phi ~.
    \label{eq:VasGradPhi0}
\end{eqnarray}
Therefore, the last term in eq.~\eqref{eq:GT_H_3} can be rewritten as
\begin{subequations}
\lb{eq:BoostTerm}
\begin{align}
    \int_{\mathcal{D}}^{} d^3\bx ~ \bV \cdot {\bomega} ~ &= \int_{\mathcal{D}}^{} d^3\bx ~ {\grad \phi} \cdot {\bomega} ~ \lb{eq:BoostTerm_1}\\
    &= \int_{\mathcal{D}}^{} d^3\bx ~ \grad \cdot (\phi \,\bomega) ~ \lb{eq:BoostTerm_2}\\
    &= {\int_{\mathcal{S}} \phi\, \bomega \cdot \bnh~dS }~,\lb{eq:BoostTerm_3}
\end{align}
\end{subequations}
where $dS$ is a surface element, and $\bnh$ is again the local normal vector to $\cS$. We used $\grad\bdot\bomega=0$ to arrive at eq.~\eqref{eq:BoostTerm_2} and Gauss's theorem to derive the last expression.

From the designation of region $\cD$ in eq.~\eqref{eq:meanHelicity} as being enclosed by a vorticity surface,
\begin{eqnarray}
    \bomega \cdot \bnh = 0 \hspace{.5cm} \mbox{over} \hspace{.5cm} \cS,
    \label{eq:omegaDotn}
\end{eqnarray}
expression~\eqref{eq:BoostTerm_3} vanishes, yielding

\begin{eqnarray}
    \int_{\mathcal{D}}^{} d^3\bx ~ \bV \cdot {\bomega} ~ = 0~.
    \label{eq:R4}
\end{eqnarray}
Finally, we have from eq.~\eqref{eq:GT_H} that mean kinetic helicity is invariant under Galilean transformations, 
\begin{eqnarray}
    {\mathcal{H}}^{'} &=& \mathcal{H}~,
    \label{eq:finalResult}
\end{eqnarray}
over regions $\cD$ enclosed by vorticity surfaces.
%Therefore, it is now clear that the domain-averaged kinetic Helicity remains invariant under any Galilean Transformation.

%%%%%%%%%%%%%%%%%%%%%%%%%%%%%%%%%%%%%%%%%%
% \section{Galilean Invariance and Gauge Invariance}
Galilean invariance of mean kinetic helicity, $\cH$, can be regarded as a special case of gauge invariance of $\cH$ as defined in eq.~\eqref{eq:meanHelicity} over a region $\cD$ enclosed by a vorticity surface \cite{moffatt1969degree,berger1984topological}. The condition that $\cD$ be enclosed by a vorticity surface can be removed by considering \emph{relative kinetic helicity},
\be h_R \equiv (\bu + \bu_*)\bdot (\bomega - \bomega_*),
\lb{eq:RelativeHelicity}\ee
in analogy with relative magnetic helicity \cite{berger1984topological,FinnAntonsen1985}. It measures helicity relative to a reference field $\bomega_*=\grad\btimes\bu_*$, which is a solenoidal vector field frozen to the flow $\bu$ within any volume $\cD$ enclosed by surface $\cS$ that is not necessarily a vorticity surface:
\begin{subequations}
\lb{eq:omega*}
\begin{align}
    \partial_t \bomega_* &= \grad\btimes(\bu\btimes\bomega_*)~,  \lb{eq:omega*_1}\\
    \bomega_* &= \grad\btimes\bu_*~,  \lb{eq:omega*_2}\\
    \bomega_*\bdot\bnh|_\cS &= \bomega\bdot\bnh|_\cS ~. \lb{eq:omega*_3}
\end{align}
\end{subequations}
The vorticity flux matching condition~\eqref{eq:omega*_3} ensures that the normal component of $\bomega-\bomega_*$ vanishes at the surface $\cS$. The field $\bomega_*$ can extend beyond $\cD$. Here, $\bu_*$ is merely a vector potential,  not necessarily related to the flow velocity $\bu$. From eq.~\eqref{eq:omega*}, $\bu_*$ is governed by 
%\begin{subequations}
\be
%\lb{eq:bu*}
%\begin{align}
    \partial_t \bu_* = \bu\btimes\bomega_* - \grad\eta_*~,  \lb{eq:bu*}
%\end{align}
%\end{subequations}
\ee
where $\eta_*$ is a scalar potential analogous to pressure in the momentum equation or the electrostatic potential in the equation for vector potential of a magnetic field \cite{Aluie17}. 

Existence of a reference field $\bomega_*$ satisfying eqs.~\eqref{eq:omega*_2},\eqref{eq:omega*_3} over a smooth surface $\cS$ at any time $t_0$ can be shown as follows. Since $\bomega_*$ is divergence-free from eq.~\eqref{eq:omega*_2}, by choosing $\bomega_* = \grad \chi$, existence of $\bomega_*$ then follows from the existence of a harmonic scalar field $\chi$ satisfying Neumann boundary conditions:
\begin{subequations}
\lb{eq:Laplace}
\begin{align}
    &\nabla^2\chi = 0 \hspace{1.03cm} \mbox{over} \hspace{.5cm} \cD~,  \lb{eq:Laplace_1}\\
    &\bnh\bdot\grad\chi = \bomega\bdot\bnh \hspace{.5cm} \mbox{over} \hspace{.5cm} \cS~.  \lb{eq:Laplace_2}
\end{align}
\end{subequations}
Existence (and uniqueness) of a solution to the Laplace boundary value problem in eq.~\eqref{eq:Laplace} is guaranteed from basic theory of partial differential equations (e.g. Ref.\cite{evans2022partial}) if the following solvability condition is satisfied,
\be \int_\cS \bomega\bdot\bnh ~dS = 0~.
\lb{eq:SolvabilityCondition}
\ee
For our problem, the solvability condition is always satisfied since $\bomega$ is divergence-free,
\be \int_\cS \bomega\bdot\bnh ~dS = \int_\cD d^3\bx ~ \grad\bdot\bomega = 0~.
\lb{eq:SolvabilityCondition_2}
\ee
Since $\bomega_*$ is frozen to the flow $\bu$ by eq.~\eqref{eq:omega*_1}, if $\bomega_*$ satisfies the vorticity flux matching condition~\eqref{eq:omega*_3} at time $t_0$, it will satisfy it for all later times $t>t_0$ when $\cS=\cS(t)$ is a material surface\footnote{An implicit assumption is that the flow is sufficiently smooth, which is in any case required for conservation laws to hold\cite{Aluie17,eyink2008dissipative}.}. Note that if the vorticity field is irrotational, $\bomega=\grad \alpha$, then our choice for the reference field $\bomega_*=\grad \chi$ implies that relative helicity $h_R=0$ inside $\cD$ where $\bomega_*=\bomega$ by uniqueness of the solution to eq.~\eqref{eq:Laplace}. Having $ h_R=0$ for a vorticity field that is a potential gradient is only natural from a physical standpoint. Therefore, with the choice of $\bomega_*=\grad \chi$, we can interpret $h_R$ in eq.~\eqref{eq:RelativeHelicity} as a measure of the knottedness of vorticity field lines relative to the base state of an irrotational field.

Similar to $\cH$, mean relative kinetic helicity is an ideal flow invariant. Indeed, the budget governing $h_R$ is
% \begin{subequations}
% \lb{eq:DeltahBudget}
% \begin{align}
\be
\begin{split}
&\frac{\partial}{\partial t} h_R + \grad\bdot( h_R\, \bu)= %\hspace{5cm}\\
%\hspace{2cm}
\grad\bdot\Bigg\{ \left[(\bu+\bu_*)\bdot\bu\right](\bomega-\bomega_*) \\
 &\hspace{3cm}- \left( \frac{1}{\rho_0}P+\frac{1}{2}|\bu|^2 + \eta_* \right)(\bomega-\bomega_*) \Bigg\}~,
\end{split}  \lb{eq:DeltahBudget}
\ee
where $P$ is pressure appearing in the fluid's momentum equation and $\rho_0$ is mass density. When integrated over material volume $\cD(t)$ co-moving with the flow velocity $\bu$, divergence of the right-hand-side terms proportional to $\bomega-\bomega_*$ vanishes due to condition~\eqref{eq:omega*_3}, yielding
% \begin{subequations}
% \lb{eq:DeltaHBudget}
% \begin{align}
\be
\frac{d}{dt} \cH_R = \int_{\cD(t)} d^3\bx \left[\frac{\partial}{\partial t}  h_R + \grad\bdot(  h_R\, \bu)\right]~=0.  \lb{eq:DeltaHBudget}
\ee
% \end{align}
% \end{subequations}
Here, $\cH_R$ is mean relative kinetic helicity, 
\be \cH_R \equiv \int_{\cD} d^3\bx ~ h_R~.
\ee
We used the Reynolds transport theorem for the first equality in eq.~\eqref{eq:DeltaHBudget}.
%Eq.~\eqref{eq:DeltaHBudget} implies that the rate of change of $\cH_R$ vanishes if the net advective flux of $ h_R$  (by the flow $\bu$) out of the volume $\cD$ is zero. 
Therefore, $\cH_R$ is an ideal invariant of the flow over any material volume.

In addition to being a flow invariant, $\cH_R$ is gauge invariant \cite{berger1984topological,FinnAntonsen1985}, i.e. it is invariant to the transformation
\begin{subequations}
\lb{eq:DeltaHmap}
\begin{align}
\bu \hspace{.2cm} &\longmapsto\hspace{.2cm} \bu + \grad \phi~,\\
\bu_* \hspace{.03cm} &\longmapsto\hspace{.2cm} \bu_* + \grad \phi_*~
\end{align}
\end{subequations}
for any scalar fields $\phi$ and $\phi_*$.
Gauge invariance of $\cH_R$ can be shown as follows, 
\begin{subequations}
\lb{eq:DeltaHinvariant}
\begin{align}
\cH'_R &= \int_\cD d^3\bx ~ (\bu + \grad\phi + \bu_* + \grad\phi_*)\bdot(\bomega-\bomega_*) \\
\begin{split}
&= \int_\cD d^3\bx ~ (\bu + \bu_*)\bdot(\bomega-\bomega_*) \\
&\quad + \int_\cD d^3\bx ~ \grad\bdot\left(\left(\phi+\phi_*\right)\left(\bomega-\bomega_*\right) \right)
\end{split}\\
&= \hspace{1cm}\cH_R \hspace{1.5cm}+ \hspace{1cm}0~.
\end{align}
\end{subequations}
The second integral vanishes because of condition $(\bomega-\bomega_*)\bdot\bnh =0$ on $\cS$ from eq.~\eqref{eq:omega*_3}. A special case of the gauge transformation in eq.~\eqref{eq:DeltaHmap} is the Galilean transformation
\begin{subequations}
\lb{eq:DeltaHGalilean}
\begin{align}
\bu \hspace{.2cm} &\longmapsto\hspace{.2cm} \bu + \bV~,\\
\bu_* \hspace{.03cm} &\longmapsto\hspace{.2cm} \bu_* + \bV_*~,
\end{align}
\end{subequations}
where $\bV$ and $\bV_*$ are constant velocity boosts. Therefore, mean relative kinetic helicity, $\cH_R$, is Galilean invariant over any region in the flow.

{\bf Acknowledgement} This research was funded by US DOE grant DE-SC0020229. Partial support from US NSF grants PHY-2020249 and PHY-2206380 is acknowledged. HA was also supported by US DOE grants DE-SC0014318, DE-SC0019329, US NSF grants OCE-2123496, US NASA grant 80NSSC18K0772, and US NNSA grants DE-NA0003856, DE-NA0003914, DE-NA0004134.

\nocite{*}
\bibliography{PoF23references}% Produces the bibliography via BibTeX.

%merlin.mbs aipnum4-1.bst 2010-07-25 4.21a (PWD, AO, DPC) hacked
%Control: key (0)
%Control: author (8) initials jnrlst
%Control: editor formatted (1) identically to author
%Control: production of article title (0) allowed
%Control: page (1) range
%Control: year (1) truncated
%Control: production of eprint (0) enabled
\begin{thebibliography}{24}%
\makeatletter
\providecommand \@ifxundefined [1]{%
 \@ifx{#1\undefined}
}%
\providecommand \@ifnum [1]{%
 \ifnum #1\expandafter \@firstoftwo
 \else \expandafter \@secondoftwo
 \fi
}%
\providecommand \@ifx [1]{%
 \ifx #1\expandafter \@firstoftwo
 \else \expandafter \@secondoftwo
 \fi
}%
\providecommand \natexlab [1]{#1}%
\providecommand \enquote  [1]{``#1''}%
\providecommand \bibnamefont  [1]{#1}%
\providecommand \bibfnamefont [1]{#1}%
\providecommand \citenamefont [1]{#1}%
\providecommand \href@noop [0]{\@secondoftwo}%
\providecommand \href [0]{\begingroup \@sanitize@url \@href}%
\providecommand \@href[1]{\@@startlink{#1}\@@href}%
\providecommand \@@href[1]{\endgroup#1\@@endlink}%
\providecommand \@sanitize@url [0]{\catcode `\\12\catcode `\$12\catcode `\&12\catcode `\#12\catcode `\^12\catcode `\_12\catcode `\%12\relax}%
\providecommand \@@startlink[1]{}%
\providecommand \@@endlink[0]{}%
\providecommand \url  [0]{\begingroup\@sanitize@url \@url }%
\providecommand \@url [1]{\endgroup\@href {#1}{\urlprefix }}%
\providecommand \urlprefix  [0]{URL }%
\providecommand \Eprint [0]{\href }%
\providecommand \doibase [0]{http://dx.doi.org/}%
\providecommand \selectlanguage [0]{\@gobble}%
\providecommand \bibinfo  [0]{\@secondoftwo}%
\providecommand \bibfield  [0]{\@secondoftwo}%
\providecommand \translation [1]{[#1]}%
\providecommand \BibitemOpen [0]{}%
\providecommand \bibitemStop [0]{}%
\providecommand \bibitemNoStop [0]{.\EOS\space}%
\providecommand \EOS [0]{\spacefactor3000\relax}%
\providecommand \BibitemShut  [1]{\csname bibitem#1\endcsname}%
\let\auto@bib@innerbib\@empty
%</preamble>
\bibitem [{\citenamefont {Moreau}(1961)}]{moreau1961constants}%
  \BibitemOpen
  \bibfield  {author} {\bibinfo {author} {\bibfnamefont {J.~J.}\ \bibnamefont {Moreau}},\ }\bibfield  {title} {\enquote {\bibinfo {title} {Constants of a {\^\i}vortex batch in perfect barotropic fluid},}\ }\href@noop {} {\bibfield  {journal} {\bibinfo  {journal} {Weekly minutes of the meetings of the Academy of Sciences}\ }\textbf {\bibinfo {volume} {252}},\ \bibinfo {pages} {2810--2812} (\bibinfo {year} {1961})}\BibitemShut {NoStop}%
\bibitem [{\citenamefont {Elsasser}(1956)}]{elsasser1956hydromagnetic}%
  \BibitemOpen
  \bibfield  {author} {\bibinfo {author} {\bibfnamefont {W.~M.}\ \bibnamefont {Elsasser}},\ }\bibfield  {title} {\enquote {\bibinfo {title} {Hydromagnetic dynamo theory},}\ }\href@noop {} {\bibfield  {journal} {\bibinfo  {journal} {Reviews of modern Physics}\ }\textbf {\bibinfo {volume} {28}},\ \bibinfo {pages} {135} (\bibinfo {year} {1956})}\BibitemShut {NoStop}%
\bibitem [{\citenamefont {Woltjer}(1958)}]{woltjer1958theorem}%
  \BibitemOpen
  \bibfield  {author} {\bibinfo {author} {\bibfnamefont {L.}~\bibnamefont {Woltjer}},\ }\bibfield  {title} {\enquote {\bibinfo {title} {A theorem on force-free magnetic fields},}\ }\href@noop {} {\bibfield  {journal} {\bibinfo  {journal} {Proceedings of the National Academy of Sciences}\ }\textbf {\bibinfo {volume} {44}},\ \bibinfo {pages} {489--491} (\bibinfo {year} {1958})}\BibitemShut {NoStop}%
\bibitem [{\citenamefont {Moffatt}(1969)}]{moffatt1969degree}%
  \BibitemOpen
  \bibfield  {author} {\bibinfo {author} {\bibfnamefont {H.~K.}\ \bibnamefont {Moffatt}},\ }\bibfield  {title} {\enquote {\bibinfo {title} {The degree of knottedness of tangled vortex lines},}\ }\href@noop {} {\bibfield  {journal} {\bibinfo  {journal} {Journal of Fluid Mechanics}\ }\textbf {\bibinfo {volume} {35}},\ \bibinfo {pages} {117--129} (\bibinfo {year} {1969})}\BibitemShut {NoStop}%
\bibitem [{\citenamefont {{Arnold}}(1973)}]{Arnold1973}%
  \BibitemOpen
  \bibfield  {author} {\bibinfo {author} {\bibfnamefont {V.~I.}\ \bibnamefont {{Arnold}}},\ }\bibfield  {title} {\enquote {\bibinfo {title} {Proc. summer school in diff. equations ( dilizhan, erevan); english transl.: (1986) sel. math. sov.}}\ \ }(\bibinfo {year} {1973})\ p.\ \bibinfo {pages} {327}\BibitemShut {NoStop}%
\bibitem [{\citenamefont {Arnold}\ and\ \citenamefont {Khesin}(1992)}]{arnold1992topological}%
  \BibitemOpen
  \bibfield  {author} {\bibinfo {author} {\bibfnamefont {V.~I.}\ \bibnamefont {Arnold}}\ and\ \bibinfo {author} {\bibfnamefont {B.~A.}\ \bibnamefont {Khesin}},\ }\bibfield  {title} {\enquote {\bibinfo {title} {Topological methods in hydrodynamics},}\ }\href@noop {} {\bibfield  {journal} {\bibinfo  {journal} {Annual review of fluid mechanics}\ }\textbf {\bibinfo {volume} {24}},\ \bibinfo {pages} {145--166} (\bibinfo {year} {1992})}\BibitemShut {NoStop}%
\bibitem [{\citenamefont {Pouquet}\ \emph {et~al.}(2019)\citenamefont {Pouquet}, \citenamefont {Rosenberg}, \citenamefont {Stawarz},\ and\ \citenamefont {Marino}}]{pouquet2019helicity}%
  \BibitemOpen
  \bibfield  {author} {\bibinfo {author} {\bibfnamefont {A.}~\bibnamefont {Pouquet}}, \bibinfo {author} {\bibfnamefont {D.}~\bibnamefont {Rosenberg}}, \bibinfo {author} {\bibfnamefont {J.~E.}\ \bibnamefont {Stawarz}}, \ and\ \bibinfo {author} {\bibfnamefont {R.}~\bibnamefont {Marino}},\ }\bibfield  {title} {\enquote {\bibinfo {title} {Helicity dynamics, inverse, and bidirectional cascades in fluid and magnetohydrodynamic turbulence: a brief review},}\ }\href@noop {} {\bibfield  {journal} {\bibinfo  {journal} {Earth and Space Science}\ }\textbf {\bibinfo {volume} {6}},\ \bibinfo {pages} {351--369} (\bibinfo {year} {2019})}\BibitemShut {NoStop}%
\bibitem [{ang(2021)}]{angriman2021broken}%
  \BibitemOpen
  \bibfield  {title} {\enquote {\bibinfo {title} {Broken mirror symmetry of tracer’s trajectories in turbulence},}\ }\href@noop {} {\bibfield  {journal} {\bibinfo  {journal} {Physical review letters}\ }\textbf {\bibinfo {volume} {127}},\ \bibinfo {pages} {254502} (\bibinfo {year} {2021})}\BibitemShut {NoStop}%
\bibitem [{\citenamefont {Speziale}(1987)}]{speziale1987helicity}%
  \BibitemOpen
  \bibfield  {author} {\bibinfo {author} {\bibfnamefont {C.~G.}\ \bibnamefont {Speziale}},\ }\bibfield  {title} {\enquote {\bibinfo {title} {On helicity fluctuations in turbulence},}\ }\href@noop {} {\bibfield  {journal} {\bibinfo  {journal} {Quarterly of applied mathematics}\ }\textbf {\bibinfo {volume} {45}},\ \bibinfo {pages} {123--129} (\bibinfo {year} {1987})}\BibitemShut {NoStop}%
\bibitem [{\citenamefont {Moffatt}\ and\ \citenamefont {Tsinober}(1992)}]{moffatt1992helicity}%
  \BibitemOpen
  \bibfield  {author} {\bibinfo {author} {\bibfnamefont {H.~K.}\ \bibnamefont {Moffatt}}\ and\ \bibinfo {author} {\bibfnamefont {A.}~\bibnamefont {Tsinober}},\ }\bibfield  {title} {\enquote {\bibinfo {title} {Helicity in laminar and turbulent flow},}\ }\href@noop {} {\bibfield  {journal} {\bibinfo  {journal} {Annual review of fluid mechanics}\ }\textbf {\bibinfo {volume} {24}},\ \bibinfo {pages} {281--312} (\bibinfo {year} {1992})}\BibitemShut {NoStop}%
\bibitem [{\citenamefont {Pope}(2000)}]{pope2000turbulent}%
  \BibitemOpen
  \bibfield  {author} {\bibinfo {author} {\bibfnamefont {S.~B.}\ \bibnamefont {Pope}},\ }\href@noop {} {\emph {\bibinfo {title} {Turbulent flows}}}\ (\bibinfo  {publisher} {Cambridge university press},\ \bibinfo {year} {2000})\BibitemShut {NoStop}%
\bibitem [{\citenamefont {Biferale}, \citenamefont {Musacchio},\ and\ \citenamefont {Toschi}(2012)}]{biferale2012inverse}%
  \BibitemOpen
  \bibfield  {author} {\bibinfo {author} {\bibfnamefont {L.}~\bibnamefont {Biferale}}, \bibinfo {author} {\bibfnamefont {S.}~\bibnamefont {Musacchio}}, \ and\ \bibinfo {author} {\bibfnamefont {F.}~\bibnamefont {Toschi}},\ }\bibfield  {title} {\enquote {\bibinfo {title} {Inverse energy cascade in three-dimensional isotropic turbulence},}\ }\href@noop {} {\bibfield  {journal} {\bibinfo  {journal} {Physical review letters}\ }\textbf {\bibinfo {volume} {108}},\ \bibinfo {pages} {164501} (\bibinfo {year} {2012})}\BibitemShut {NoStop}%
\bibitem [{\citenamefont {Baj}, \citenamefont {Portela},\ and\ \citenamefont {Carter}(2022)}]{baj2022simultaneous}%
  \BibitemOpen
  \bibfield  {author} {\bibinfo {author} {\bibfnamefont {P.}~\bibnamefont {Baj}}, \bibinfo {author} {\bibfnamefont {F.~A.}\ \bibnamefont {Portela}}, \ and\ \bibinfo {author} {\bibfnamefont {D.}~\bibnamefont {Carter}},\ }\bibfield  {title} {\enquote {\bibinfo {title} {On the simultaneous cascades of energy, helicity, and enstrophy in incompressible homogeneous turbulence},}\ }\href@noop {} {\bibfield  {journal} {\bibinfo  {journal} {Journal of Fluid Mechanics}\ }\textbf {\bibinfo {volume} {952}},\ \bibinfo {pages} {A20} (\bibinfo {year} {2022})}\BibitemShut {NoStop}%
\bibitem [{\citenamefont {Frisch}, \citenamefont {She},\ and\ \citenamefont {Sulem}(1987)}]{frisch1987large}%
  \BibitemOpen
  \bibfield  {author} {\bibinfo {author} {\bibfnamefont {U.}~\bibnamefont {Frisch}}, \bibinfo {author} {\bibfnamefont {Z.~S.}\ \bibnamefont {She}}, \ and\ \bibinfo {author} {\bibfnamefont {P.}~\bibnamefont {Sulem}},\ }\bibfield  {title} {\enquote {\bibinfo {title} {Large-scale flow driven by the anisotropic kinetic alpha effect},}\ }\href@noop {} {\bibfield  {journal} {\bibinfo  {journal} {Physica D: Nonlinear Phenomena}\ }\textbf {\bibinfo {volume} {28}},\ \bibinfo {pages} {382--392} (\bibinfo {year} {1987})}\BibitemShut {NoStop}%
\bibitem [{\citenamefont {Inagaki}, \citenamefont {Yokoi},\ and\ \citenamefont {Hamba}(2017)}]{inagaki2017mechanism}%
  \BibitemOpen
  \bibfield  {author} {\bibinfo {author} {\bibfnamefont {K.}~\bibnamefont {Inagaki}}, \bibinfo {author} {\bibfnamefont {N.}~\bibnamefont {Yokoi}}, \ and\ \bibinfo {author} {\bibfnamefont {F.}~\bibnamefont {Hamba}},\ }\bibfield  {title} {\enquote {\bibinfo {title} {Mechanism of mean flow generation in rotating turbulence through inhomogeneous helicity},}\ }\href@noop {} {\bibfield  {journal} {\bibinfo  {journal} {Physical Review Fluids}\ }\textbf {\bibinfo {volume} {2}},\ \bibinfo {pages} {114605} (\bibinfo {year} {2017})}\BibitemShut {NoStop}%
\bibitem [{\citenamefont {Teitelbaum}\ and\ \citenamefont {Mininni}(2011)}]{teitelbaum2011decay}%
  \BibitemOpen
  \bibfield  {author} {\bibinfo {author} {\bibfnamefont {T.}~\bibnamefont {Teitelbaum}}\ and\ \bibinfo {author} {\bibfnamefont {P.~D.}\ \bibnamefont {Mininni}},\ }\bibfield  {title} {\enquote {\bibinfo {title} {The decay of turbulence in rotating flows},}\ }\href@noop {} {\bibfield  {journal} {\bibinfo  {journal} {Physics of Fluids}\ }\textbf {\bibinfo {volume} {23}} (\bibinfo {year} {2011})}\BibitemShut {NoStop}%
\bibitem [{\citenamefont {Rorai}\ \emph {et~al.}(2013)\citenamefont {Rorai}, \citenamefont {Rosenberg}, \citenamefont {Pouquet},\ and\ \citenamefont {Mininni}}]{rorai2013helicity}%
  \BibitemOpen
  \bibfield  {author} {\bibinfo {author} {\bibfnamefont {C.}~\bibnamefont {Rorai}}, \bibinfo {author} {\bibfnamefont {D.}~\bibnamefont {Rosenberg}}, \bibinfo {author} {\bibfnamefont {A.}~\bibnamefont {Pouquet}}, \ and\ \bibinfo {author} {\bibfnamefont {P.~D.}\ \bibnamefont {Mininni}},\ }\bibfield  {title} {\enquote {\bibinfo {title} {Helicity dynamics in stratified turbulence in the absence of forcing},}\ }\href@noop {} {\bibfield  {journal} {\bibinfo  {journal} {Physical Review E}\ }\textbf {\bibinfo {volume} {87}},\ \bibinfo {pages} {063007} (\bibinfo {year} {2013})}\BibitemShut {NoStop}%
\bibitem [{\citenamefont {Buzzicotti}\ \emph {et~al.}(2018)\citenamefont {Buzzicotti}, \citenamefont {Aluie}, \citenamefont {Biferale},\ and\ \citenamefont {Linkmann}}]{buzzicotti2018energy}%
  \BibitemOpen
  \bibfield  {author} {\bibinfo {author} {\bibfnamefont {M.}~\bibnamefont {Buzzicotti}}, \bibinfo {author} {\bibfnamefont {H.}~\bibnamefont {Aluie}}, \bibinfo {author} {\bibfnamefont {L.}~\bibnamefont {Biferale}}, \ and\ \bibinfo {author} {\bibfnamefont {M.}~\bibnamefont {Linkmann}},\ }\bibfield  {title} {\enquote {\bibinfo {title} {Energy transfer in turbulence under rotation},}\ }\href@noop {} {\bibfield  {journal} {\bibinfo  {journal} {Physical Review Fluids}\ }\textbf {\bibinfo {volume} {3}},\ \bibinfo {pages} {034802} (\bibinfo {year} {2018})}\BibitemShut {NoStop}%
\bibitem [{\citenamefont {Berger}\ and\ \citenamefont {Field}(1984)}]{berger1984topological}%
  \BibitemOpen
  \bibfield  {author} {\bibinfo {author} {\bibfnamefont {M.~A.}\ \bibnamefont {Berger}}\ and\ \bibinfo {author} {\bibfnamefont {G.~B.}\ \bibnamefont {Field}},\ }\bibfield  {title} {\enquote {\bibinfo {title} {The topological properties of magnetic helicity},}\ }\href@noop {} {\bibfield  {journal} {\bibinfo  {journal} {Journal of Fluid Mechanics}\ }\textbf {\bibinfo {volume} {147}},\ \bibinfo {pages} {133--148} (\bibinfo {year} {1984})}\BibitemShut {NoStop}%
\bibitem [{\citenamefont {{Finn}}\ and\ \citenamefont {{Antonsen}}(1985)}]{FinnAntonsen1985}%
  \BibitemOpen
  \bibfield  {author} {\bibinfo {author} {\bibfnamefont {J.~M.}\ \bibnamefont {{Finn}}}\ and\ \bibinfo {author} {\bibfnamefont {T.~M.}\ \bibnamefont {{Antonsen}}},\ }\bibfield  {title} {\enquote {\bibinfo {title} {{Magnetic helicity: What is it and what is it good for?}}}\ }\href {\doibase 10.1016/j.physd.2006.08.009} {\bibfield  {journal} {\bibinfo  {journal} {Comments Plasma Phys. Contr. Fusion}\ }\textbf {\bibinfo {volume} {9}},\ \bibinfo {pages} {111--126} (\bibinfo {year} {1985})}\BibitemShut {NoStop}%
\bibitem [{\citenamefont {Aluie}(2017)}]{Aluie17}%
  \BibitemOpen
  \bibfield  {author} {\bibinfo {author} {\bibfnamefont {H.}~\bibnamefont {Aluie}},\ }\bibfield  {title} {\enquote {\bibinfo {title} {{Coarse-grained incompressible magnetohydrodynamics: analyzing the turbulent cascades}},}\ }\href {\doibase 10.1088/1367-2630/aa5d2f} {\bibfield  {journal} {\bibinfo  {journal} {New Journal of Physics}\ }\textbf {\bibinfo {volume} {19}},\ \bibinfo {pages} {025008} (\bibinfo {year} {2017})}\BibitemShut {NoStop}%
\bibitem [{\citenamefont {Evans}(2022)}]{evans2022partial}%
  \BibitemOpen
  \bibfield  {author} {\bibinfo {author} {\bibfnamefont {L.~C.}\ \bibnamefont {Evans}},\ }\href@noop {} {\emph {\bibinfo {title} {Partial differential equations}}},\ Vol.~\bibinfo {volume} {19}\ (\bibinfo  {publisher} {American Mathematical Society},\ \bibinfo {year} {2022})\BibitemShut {NoStop}%
\bibitem [{Note1()}]{Note1}%
  \BibitemOpen
  \bibinfo {note} {An implicit assumption is that the flow is sufficiently smooth, which is in any case required for conservation laws to hold\cite {Aluie17,eyink2008dissipative}.}\BibitemShut {Stop}%
\bibitem [{\citenamefont {Eyink}(2008)}]{eyink2008dissipative}%
  \BibitemOpen
  \bibfield  {author} {\bibinfo {author} {\bibfnamefont {G.~L.}\ \bibnamefont {Eyink}},\ }\bibfield  {title} {\enquote {\bibinfo {title} {Dissipative anomalies in singular euler flows},}\ }\href@noop {} {\bibfield  {journal} {\bibinfo  {journal} {Physica D: Nonlinear Phenomena}\ }\textbf {\bibinfo {volume} {237}},\ \bibinfo {pages} {1956--1968} (\bibinfo {year} {2008})}\BibitemShut {NoStop}%
\end{thebibliography}%

\end{document}